\documentclass[aps,pre,reprint,groupedaddress,showkeys,sort&compress,superscriptaddress,nofootinbib]{revtex4-1}

\usepackage{bm}
\usepackage{epsfig}
\usepackage{rotating}
\usepackage{subfigure}
\usepackage{url}
\usepackage[nice]{nicefrac}
\usepackage{graphicx,color}
\usepackage{amsmath,amssymb,bm,bbm}
\usepackage{braket}
\usepackage{bbold}
\usepackage{float}
\usepackage{placeins}
\usepackage{soul}
\usepackage{enumitem}
\usepackage{media9}
\usepackage{booktabs}
\usepackage{longtable}
\usepackage[bottom]{footmisc}

\usepackage{breakurl}
\usepackage[breaklinks]{hyperref}

\def\be{\begin{equation}}

\def\ee{\end{equation}}
\def\barr{\begin{array}}
\def\earr{\end{array}}

\def\l{\left}
\def\r{\right}

\def\ed{\end{document}}

\begin{document}


\title{Coarse graining correlation matrices according to macrostructures: \\ Financial markets as a paradigm}


\author{M. Mija{\'i}l Mart{\'i}nez-Ramos}
\affiliation{Instituto de Ciencias F{\'i}sicas, Universidad Nacional Aut{\'o}noma de M\'{e}xico, 62210 Cuernavaca, M\'{e}xico}
\author{Parisa Majari}
\affiliation{Instituto de Ciencias F{\'i}sicas, Universidad Nacional Aut{\'o}noma de M\'{e}xico, 62210 Cuernavaca, M\'{e}xico}
\author{Andres R. Cruz-Hern{\'a}ndez}
\address{Instituto de Ciencias F{\'i}sicas, Universidad Nacional Aut{\'o}noma de M\'{e}xico, 62210 Cuernavaca, M\'{e}xico}
\address{School of Management, Universidad de los Andes, 111711,  Bogot{\'a}, Colombia}
\author{Hirdesh K.  Pharasi}
\affiliation{School of Engineering and Technology,  BML Munjal University,  Gurugram,  Haryana  122413,  India}
\author{Manan Vyas}
\email{Corresponding author; manan@icf.unam.mx}
\affiliation{Instituto de Ciencias F{\'i}sicas, Universidad Nacional Aut{\'o}noma de M\'{e}xico, 62210 Cuernavaca, M\'{e}xico}


\date{\today}

\begin{abstract}

We analyze correlation structures in financial markets by coarse graining the Pearson correlation matrices according to market sectors to obtain Guhr matrices using Guhr's correlation method according to Ref.  [P. Rinn {\it et. al.}, Europhysics Letters 110, 68003 (2015)]. We compare the results for the evolution of market states and the  corresponding transition matrices with those obtained using Pearson correlation matrices.  The behavior of market states is found to be similar for both the coarse grained and Pearson matrices.  However,  the number of relevant variables is reduced by orders of magnitude.

\end{abstract}


\maketitle

\section{Introduction}

Easy availability of financial markets data makes these ideal to explore new aspects of complex systems.  About 25 years ago, it was recognized that the state of a financial market is largely determined by largest eigenvalue of the correlation matrix of returns of the closing prices \cite{MSBook}.  In 2012, the concept of discrete `market states' was introduced based on a clustering algorithm applied in the space of correlation matrices \cite{SciRep2012} and has received considerable attention in and outside financial markets \cite{ Zhou_2018, PhysRevE.97.052312, Tang_2018, NJP2018, Springer2019, PhyA24,  Wang_2020, Nie_2020, Heckens_2020,  JAMES-2022,  JSM2022, PhyA2022}.  Note that the concept of `market states' is different from `state of the market',  which is defined by the largest eigenvalue \cite{SciRep2012}.  This concept is one option of what is often referred to as `market regimes' by economists \cite{Heckens_2020, Campbell_1997}, but the latter may use a wider range of parameters.  The correlation matrix has proven valuable in characterizing various aspects of stock returns \cite{Mant-99, Bon-01}, stock index returns and global stock returns \cite{Bon-00}, and stock return volatility \cite{Mi-03}. Precisely estimating the properties of correlation matrices is crucial for a range of financial decision-making processes such as asset allocation, portfolio optimization \cite{To-08}, and derivative pricing \cite{Tu-10}.

In Refs.  \cite{NJP2018,PhyA24,Springer2019},  it was found that a $k$-means (in what follows ``KM") clustering calculation produces results that have a marked dependence on the average correlations or the largest eigenvalue of the Pearson correlation matrix as can be expected from the standard literature \cite{MSBook,  bouchaud2003theory} \footnote{ Indeed the correlation of these two quantities over the time horizon from 2006 to 2023 is above 0.9 \cite{arXiv2023}}.  Yet we see clear differences in the S\&P 500 data for similar average correlations in two and three-dimensional representations obtained by dimensional scaling \cite{seber2009multivariate,Springer2019, NJP2018, PhyA24,  Chap23}. These results produce transition matrices between states that have given a very reasonable account of states where a crash may be imminent \cite{PhyA24}.  It is not far fetched to suspect that correlations of sectors could give similar results with smaller effort and possibly even clearer signals.

The simple-minded approach of averaging over the returns of stocks in each sector is not satisfactory for a variety of reasons. Fortunately in Ref.  \cite{Rinn_2015} a coarse-graining (in what follows ``CG") of the correlation was proposed, where we average correlation matrix elements over sectorial blocks providing sort of a covariance matrix over the market sectors. This matrix, which we shall call the `sectorial Guhr matrix', provides the appropriate tool for our endeavor.

In the present paper, we shall analyze S$\&$P 500 and Nikkei 225 using the stocks that exist over our time horizon, which we choose identical to Refs. \cite{PhyA24,Chap23}. We will thus repeat the analysis of Ref \cite{PhyA24} in some detail and compare the results of noise reduction by a power map (hereinafter ``PM") \cite{Guhr_2003, Guhr_2010, THS_2013},  and KM \cite{teknomo2006k,jain2010data} with the corresponding results for the sectorial Guhr matrices.  We also plot the dimensionally scaled matrices for visualization purposes.  Indeed, if we think of parametrizing the correlation matrix, for example if we have $N = 350$ stocks for S$\&$P 500 market (after purification of the time series), we obtain 350 time series and thus a $350\times350$ correlation matrix: due to the symmetry of the matrix and the diagonal consisting of ones we have 61075 parameters [$N(N-1)/2$ parameters], which is not very pleasant. The $10 \times 10$ sectorial Guhr matrix for market sectors will have $55$ parameters, which is not great but provides a dramatic reduction. 

In the next section, we shall detail the techniques and describe the data sets we use, following ideas outlined in \cite{NJP2018, Springer2019, PhyA24}.  In section \ref{sec1}, we will explain the technique of Guhr et. al.  \cite{Rinn_2015} to obtain the sectorial Guhr matrices from the Pearson correlation matrices.  We shall construct the correlation matrices as a function of the trading days with epoch lengths of 20 and one day shift of the epoch for the daily closing price returns.  Note that the closing prices data are used as some random walk properties are expected for intra-day trading and opening data seem to be more volatile than closing data; also they don’t reflect the same degree of the decisions of investors/investment groups that are based on fundamental data. We will then apply the clustering technique.  In section \ref{sec2}, we show the corresponding results for S\&P 500 and Nikkei 225 with time horizon of January 2006 to December 2019.  Note that we exclude COVID period as there is a significant change in the structure of market states observed recently \cite{arXiv2023}, with an entirely new state appearing. Yet only time can tell, if this is a change in the structure of the market or a transitory single occurrence that is tapering off.  We finally will present conclusions and an outlook of possible extensions in the methods and in the applications.

\section{Data and techniques used}
\label{sec1}

We use the daily adjusted closing price $P_i=[p_1,p_2,...,p_{T}]$  of $i$-th stock of S$\&$P 500  (USA) and Nikkei 225 indices over the period from January 2006 to December 2019.  Here $i=1,2,...,N$; N denotes the number of stocks and $T$ is the time horizon. The S$\&$P 500  market consists of $N = 350$ number of stocks over $T= 3523$ trading days and the Nikkei 225 consists of $N = 150$ stocks over $T = 3458 $ trading days.  Note that here we choose the same time horizon as Ref.  \cite{PhyA24}.  Note that $N$ is usually smaller than the number of listed stocks as we use the criterion to exclude stocks that have not been traded for more than two consecutive days during the chosen time horizon.  

Using the time series for $P_i$, we compute time series for logarithmic returns $r_i(t) = \ln p_i(t+\Delta t)-\ln p_i(t)$; $t = 1,  2,  \ldots, T-1$.  Then, the Pearson correlation $C_{ij}(\tau)$ between stocks $i$ and $j$ is  defined as 
\begin{equation}
C_{ij}(\tau)={[r_i- \langle r_i \rangle][r_j - \langle r_j \rangle]}/{\sqrt{\mbox{var}(r_i) \; \mbox{var}(r_j)}} \;;
\label{eq-corr}
\end{equation} 
here, the symbol $\langle \ldots \rangle$ denotes average and var$(\ldots)$ is the variance of the return time series for each epoch:  $\mbox{var}(r_i) = \{\sum_{j = \tau_t}^{\tau_{t^\prime}} r_{i,j}\}/(\tau_{t^\prime} - \tau_t +1)$.  In Eq. \eqref{eq-corr},  $\tau$ is the end date of the epoch which we shall choose as $20$ trading days, i.e. approximately one month. To obtain the time evolution we shift the epochs by one day over the time horizon.  Thus, we have $M = 3503$ and $3438$ correlations matrices for S$\&$P 500 and Nikkei 225, respectively.

\begin{table}[!htb]
\caption{Sectors in S\&P 500 market}
\begin{tabular}{ccccc}
    \toprule
    \textbf{Label} & \textbf{Sectors}  \\
    \midrule
    CD & Consumer Discretionary \\ \hline
CS & Consumer Staples \\ \hline
EG & Energy\\ \hline
FN  & Financials \\ \hline
HC& Health Care \\ \hline
ID& Industrials \\ \hline
IT& Information Technology \\ \hline
MT  & Materials \\ \hline
TC & Technology \\ \hline
UT & Utilities \\ \hline
    \midrule
\end{tabular}    
\label{tab:1}
\end{table}

\begin{table}[!htb]
\caption{Sectors in Nikkei 225 market}
\begin{tabular}{ccccc}
    \toprule
    \textbf{Label} & \textbf{Sectors}  \\
    \midrule
   CG &Consumer Goods \\ \hline
CP & Capital Goods/Others \\ \hline
FN  & Financials \\ \hline
MT  & Materials\\ \hline
TC & Technology \\ \hline
UT & Transportation $\&$ Utilities \\ \hline
    \midrule
\end{tabular}    
\label{tab:2}
\end{table}

We apply the PM method as a noise suppression technique \cite{Guhr_2003, Guhr_2010, THS_2013} on Pearson correlation matrices using the following definition:
\begin{equation}\label{4}
C^\prime_{ij}(\tau)=sign(C_{ij})|C_{ij}|^{1+\epsilon}
\end{equation}
where $0< \epsilon <1$  is the noise-suppression parameter.  Note that the applying PM method on the CG matrices gives similar results as if we apply PM method on the correlation matrices and then obtain the CG matrices. 

\begin{figure*}
            \subfigure[]{\includegraphics[width=8.6cm]{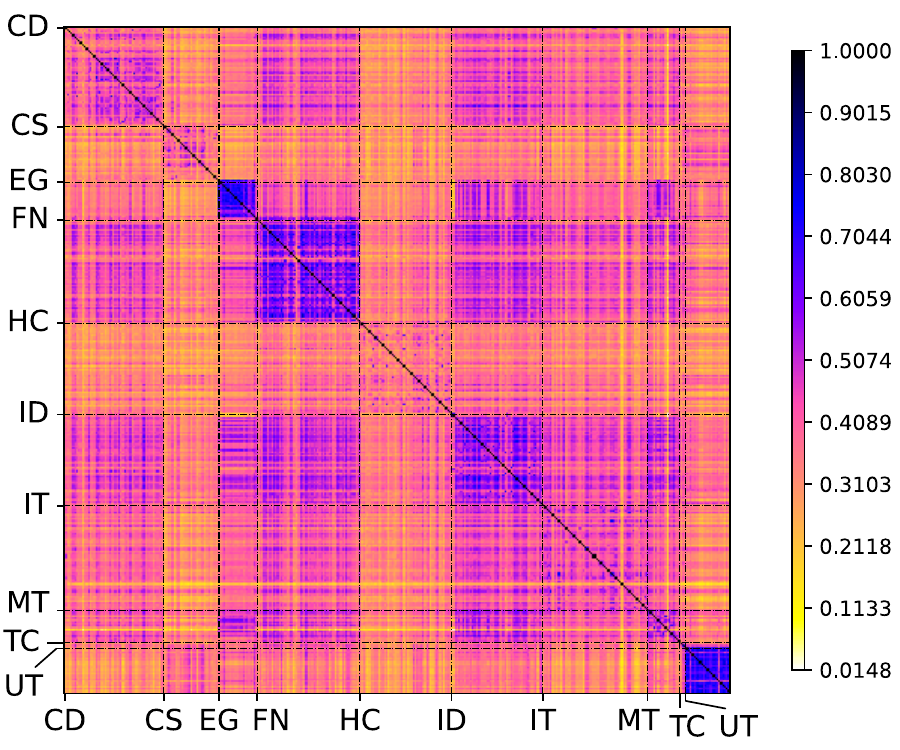}\label{fig:1a}}
             \subfigure[]{\includegraphics[width=8.6cm]{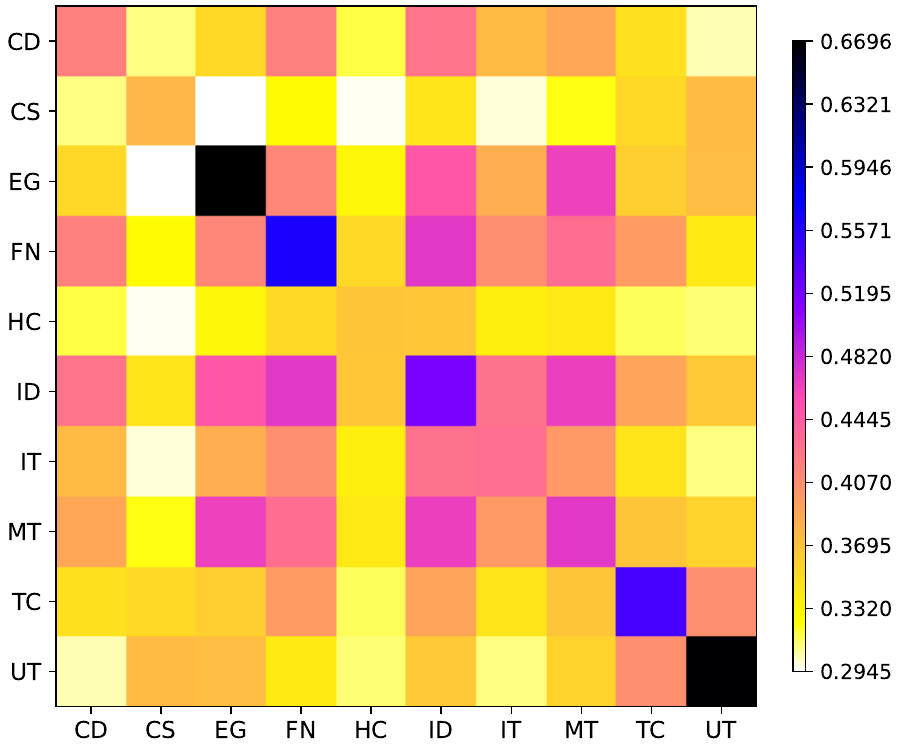}\label{fig:1b}}
\caption{Correlation matrix plot for the total time horizon considered (January 2006 to December 2019) with (a) Pearson correlation and (b) sectorial Guhr matrix for S\&P 500. The market sectors (see Table \ref{tab:1}) for the Guhr matrices are not scaled according to the number of stocks in the particular sector. Note that no negative correlations survive over the total time horizon.}
\label{fig:1}
\end{figure*}

\begin{figure*}
            \subfigure[]{\includegraphics[width=8.6cm]{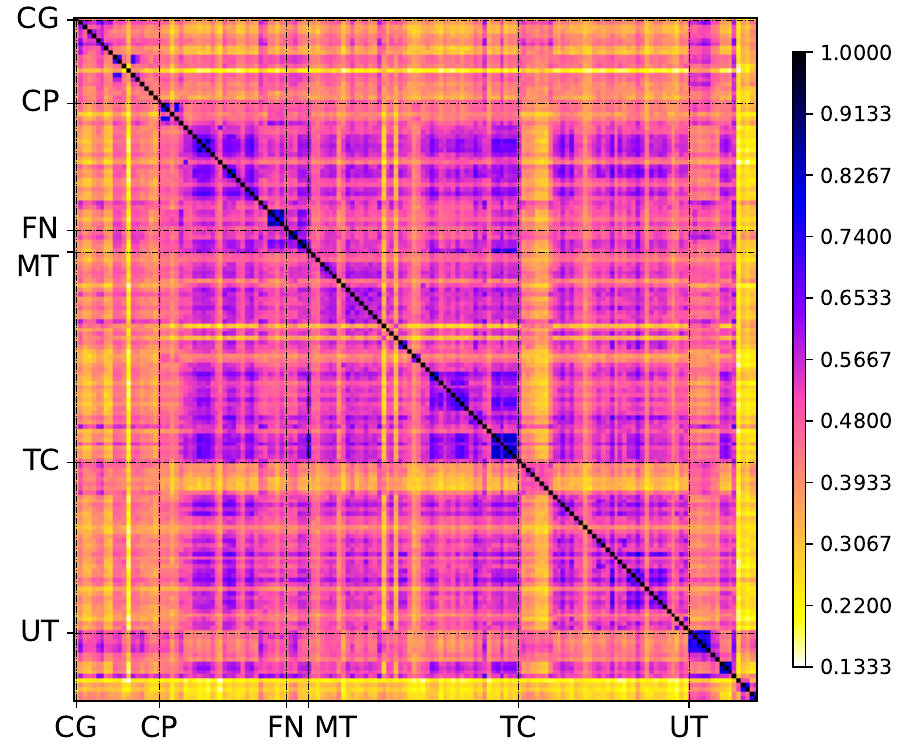}\label{fig:2a}}
             \subfigure[]{\includegraphics[width=8.6cm]{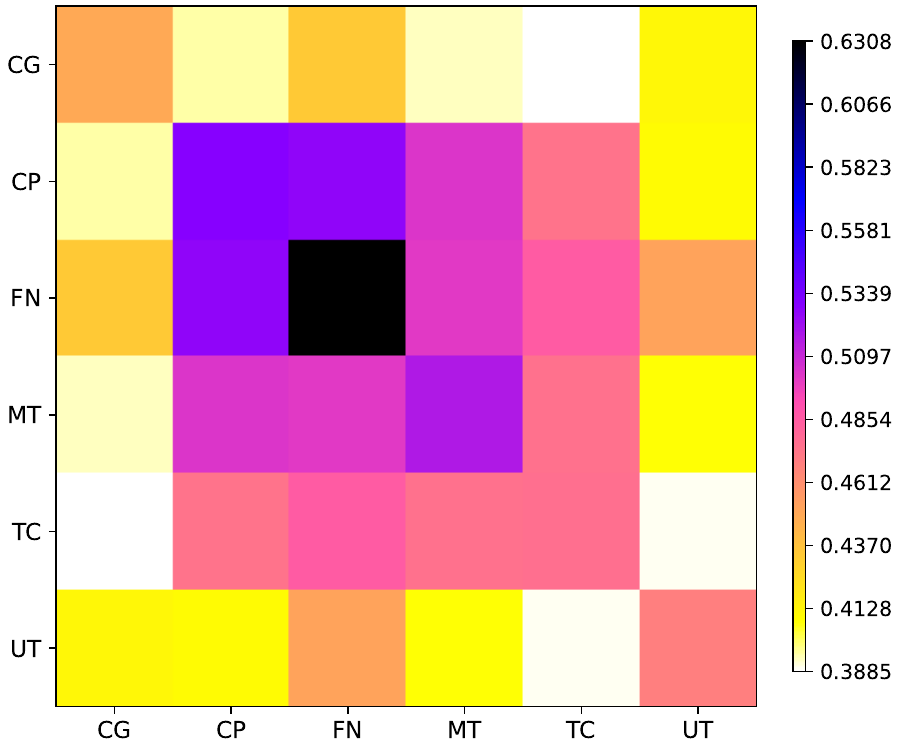}\label{fig:2b}}
\caption{Same as Fig. \ref{fig:1} but for Nikkei 225. The market sectors (see Table \ref{tab:2}) for the Guhr matrices are not scaled according to the number of stocks in the particular sector. }
\label{fig:2}
\end{figure*}

The stocks in the S\&P 500  and Nikkei 225 markets are classified into $N_s = 10$ and $6$ sectors, respectively,  as shown in tables \ref{tab:1} and \ref{tab:2}.  We then obtain the ${N_s\times N_s}$ Guhr matrices by coarse graining Pearson correlation matrices taking the average over different intra- and inter-sectorial correlation matrix elements. Considering each intra- and inter-sectorial blocks as shown in Figs. \ref{fig:1a} and \ref{fig:2a},  we sum all the correlation matrix elements inside the block (excluding the self-correlations) and divide by the total number of correlation matrix elements $m$ in the block, 
\begin{equation}
G_{ij} = \displaystyle\frac{1}{m} \displaystyle\sum_{\alpha,\beta} C_{\alpha, \beta} \;\;\;\;\;\; \forall \;\; (\alpha \in i,  \beta \in j) \;.
\label{eq-sect}
\end{equation}
Note that the Guhr matrix $G$ is not a correlation matrix as its diagonal matrix elements are not identity.  An important feature of the Guhr matrices is that we deal with low dimensional space $N_s(N_s+1)/2$, i.e., $55 \; D$ for S$\&$P 500,  and $21 \; D$ for Nikkei 225 markets which is much smaller than the original $N(N-1)/2 \;D$ one,  which is not great but provides a dramatic reduction.  Figure \ref{fig:1} shows the correlation matrix plot for (a) Pearson and (b) Guhr matrices for S\&P 500 market over the total time horizon.  Similarly, Fig. \ref{fig:2} shows the correlation matrix plot for (a) Pearson and (b) Guhr matrices for Nikkei 225 market over the total time horizon.  Note that the sectors for the Guhr matrices are not scaled according to the number of stocks in the particular sector.  There are strong sectorial correlations in Energy and Utilities sectors in S\&P 500 markets and Financials sector in Nikkei 225 market. 

We know that the average correlation (or equivalently the largest eigenvalue) dominates the behavior of the correlation matrix as a function of time \cite{SciRep2012}. Thus, although we move in a space of $N \times (N-1) /2$ dimensions,  a single dimension dominates the evolution except for details.  These very high dimensions seem meaningless in view of the dominance of the average correlation and are very unwieldy if we follow,  say $N$,  stocks.  For illustration purposes,  DS was used in \cite{seber2009multivariate,  Springer2019}, but the pictures show the significance of the average correlation. We use dimensional scaling that enables us to visualize objects (set of correlation matrices) from a higher dimensional space to a low dimensional space \cite{seber2009multivariate}. The objects in the higher dimensional space satisfy
the distance/similarity between correlation matrices at different times $\tau_t$ and $\tau_t^\prime$ and are defined as 
\begin{equation}
\xi(\tau_t,\tau_t^\prime)=  \sum_{i < j} \l|C_{ij}(\tau_t)-C_{ij}(\tau_t^\prime)\r|  \;.
\end{equation}

For completeness, we optimize the number of market states as a function of power map exponent $\epsilon$ based on the minimum standard deviation of intra-cluster distances $\sigma_{intra}$.  The results are shown in Appendix \ref{app1} as we do not find any significant differences with PM. 

\section{Results and discussion}
\label{sec2}

In this section, we first present the time evolution of market states for five clusters by optimizing standard deviation of  intra-cluster distance $\sigma_{intra}$ without applying the PM and compare it to sectorial Guhr matrices resulting from the same correlation matrices.  For the analysis, we first compute the Pearson correlation matrix using return time series and then apply PM followed by CG to obtain the Guhr matrices for each epoch.  
{Remember that the ${N_s\times N_s}$ Guhr matrices are obtained by coarse graining Pearson correlation matrices taking the average over different intra- and inter-sectorial correlation matrix elements. Considering each intra- and inter-sectorial blocks as shown in Figs. \ref{fig:1a} and \ref{fig:2a},  we sum all the correlation matrix elements inside the block (excluding the self-correlations) and divide by the total number of correlation matrix elements $m$ in the block; see Eq. \eqref{eq-sect}}. Then, using the Pearson and Guhr matrices, we do KM to obtain the market states.  We also dimensionally scale the correlation matrices to three dimensions and show the 2$D$ projections. We shall do this first for S\&P 500 and then for Nikkei 225. 

\subsection{Standard and Poors 500}
\label{sp500}

\begin{figure*}
            \subfigure[Pearson correlation matrices]{\includegraphics[width=17.2cm]
{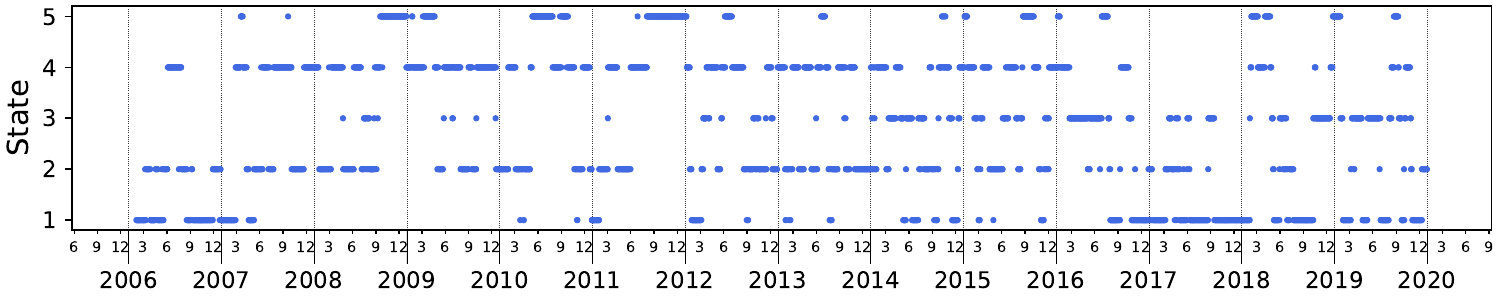}}
            \subfigure[sectorial Guhr matrices]{\includegraphics[width=17.2cm]
{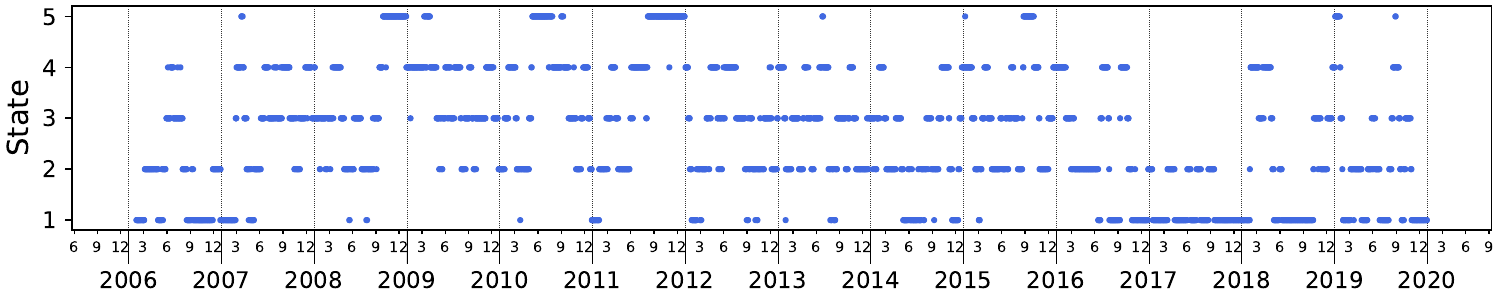}}
 \caption{Time evolution of five market states of the S\&P 500 data using (a) Pearson correlation matrices $C$ and (b) sectorial Guhr matrices $G$.  The state evolution is obtained after performing KM clustering on 3503 short-time correlation matrices.  Pearson correlation matrix elements are computed using logarithmic return time series of adjusted closing prices with epochs of length 20 trading days shifted by one trading day. The market states are arranged in order of increasing average correlations: (a) (0.157, 0.281, 0.286, 0.433, 0.611) and (b) (0.160,  0.269,  0.373,  0.487,  0.654),  respectively.}
\label{fig:3}
\end{figure*}

\begin{figure*}
            \subfigure[$\epsilon = 0.0$, Pearson correlation matrices]{\includegraphics[width=8.6cm]
{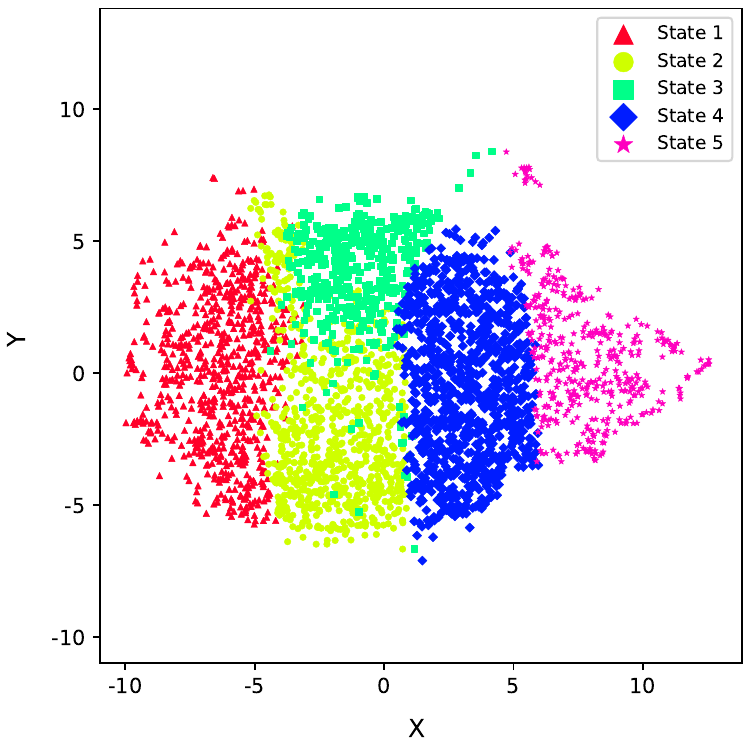}}
           \subfigure[$\epsilon = 0.0$, sectorial Guhr matrices]{\includegraphics[width=8.6cm]
{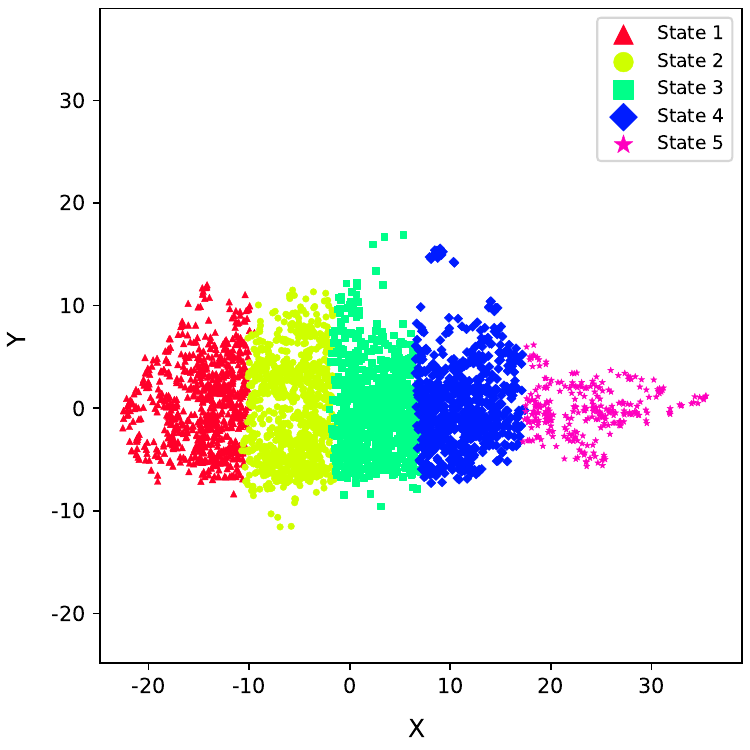}}
 \caption{3D dimensional scaling of 3503 correlation matrices of the five S\&P 500 market states from January 2006 to December 2019 shown in Fig. \ref{fig:3}: (a) $\epsilon = 0.0$, Pearson correlation matrices and (b) $\epsilon = 0.0$,  sectorial Guhr matrices.}
\label{fig:4}
\end{figure*}

\begin{figure*}[t]
    \centering
 \subfigure[Pearson correlation matrices]{\includegraphics[width=8.6cm]{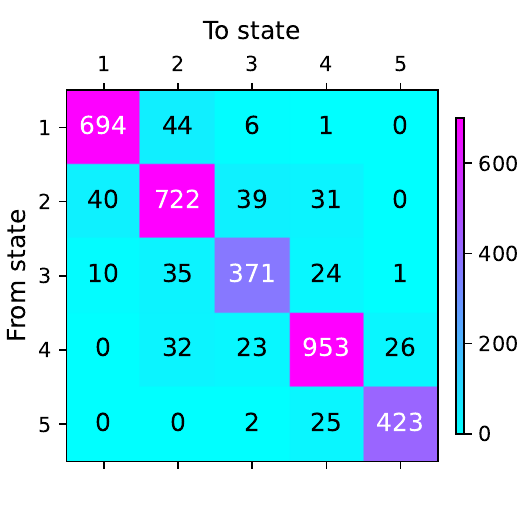}}
 \subfigure[sectorial Guhr matrices]{\includegraphics[width=8.6cm]{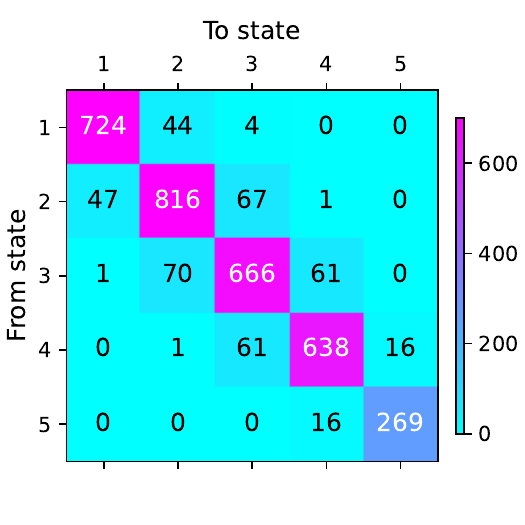}}
  \caption{Transition matrices showing transitions between different market states for S\&P 500 in Fig.  \ref{fig:3} obtained with (a) Pearson correlation matrices and (b) sectorial Guhr matrices.  The transition matrix corresponding to  Guhr matrices is near-tridiagonal.  Also, the necessary criterion for Markovianity given in Eq. (2) of \cite{NJP2018} is fulfilled for both. The equilibrium distributions corresponding to (a) and (b) are (0.2095, 0.2388, 0.1263, 0.2964, 0.129) and (0.2204, 0.2658, 0.2279, 0.2045, 0.0814) respectively.}
    \label{fig:5}
\end{figure*}

We begin our study by analyzing evolution of market states for Pearson correlation matrices and sectorial Guhr matrices as shown in Fig. \ref{fig:3} for S\&P 500.  The state evolution is obtained after performing KM clustering on the 3503 short-time correlation matrices.  Pearson correlation matrix elements are computed using logarithmic return time series of adjusted closing prices with epochs of length 20 trading days shifted by one trading day. The market states are arranged in order of increasing average correlations.  State 3 obtained using Pearson correlation matrices is markedly different from the state 3 obtained using sectorial Guhr matrices.  This difference is also reflected in the dimensional scaling figures shown in Fig. \ref{fig:4}. Also, the transition matrices are different as shown in Fig. \ref{fig:5}. Only the transition matrix corresponding to sectorial Guhr matrices is near-tridiagonal.  Also, the necessary criterion for Markovianity given in Eq. (2) of \cite{NJP2018} is fulfilled for both the transition matrices.  Using CG, the number of transitions from state 4 (near-critical state) to state 5 (critical state) reduce considerably as can be seen from Fig. \ref{fig:5}. For the sectorial Guhr matrices, there are no transitions from states 1-3 to state 5.  However, when we use PM (see Fig. \ref{fig:10} in Appendix \ref{app1}), state 3 is almost identical for both Pearson correlation matrices and sectorial Guhr matrices.  As a consequence, the dimensionally scaled figure and transition matrices for the two as shown in Figs. \ref{fig:11} and \ref{fig:12} are also very similar.  Note that dimensional scaling shows no state overlaps and transition matrices are near-tridiagonal for both.  Thus, CG does a similar job as PM and hence,  PM along with CG is same as CG.  Note that typical state concentrations are seen in Fig. \ref{fig:3} on higher side of correlations at the end of years 2008 and 2011 while the opposite happens with the quiet market region of the year 2017.  In the quiet market region of 2017, the absence of states 3-5 is prominent when using sectorial Guhr matrices.  These features are also seen when we apply PM (see Fig. \ref{fig:10} in Appendix \ref{app1}).  Coarse graining and PM both reduce the number of transitions from near-critical state to the critical state. 

Note that the S\&P 500 market with the chosen parameters shows a sequential behavior of market states as a function of average correlations or equivalently,  largest eigenvalues.  With optimal PM parameters, this behavior is to be expected as a dominance of the largest eigenvalues is well-known \cite{MSBook}.  We will call this behavior lineality of market states.  This lineality has to break down if we increase the states and indeed this will happen for the S\&P 500 market if we do not apply PM. Nevertheless, in the CG picture, lineality is conserved also for six clusters.  Thus, in this CG case, transition matrix will again become near-tridiagonal.  

\subsection{Nikkei 225}
\label{nikk}

\begin{figure*}
            \subfigure[]{\includegraphics[width=17.2cm]{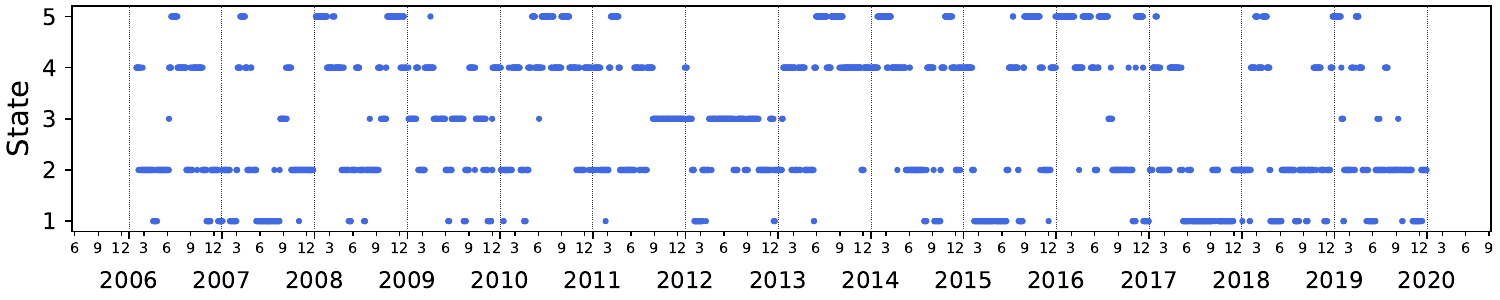}}
            \subfigure[]{\includegraphics[width=17.2cm]{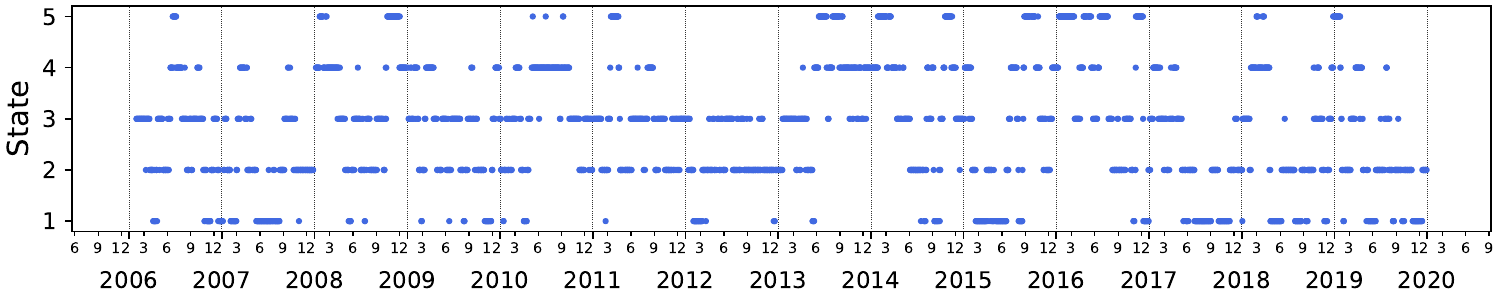}}
 \caption{Time evolution of five market states of the Nikkei 225 data using (a) Pearson correlation matrices $C$ and (b) sectorial Guhr matrices $G$.  The state evolution is obtained after performing KM clustering on 3438 short-time correlation matrices.  Pearson correlation matrix elements are computed using logarithmic return time series of adjusted closing prices with epochs of length 20 trading days shifted by one trading day.  The market states are arranged in order of increasing average correlations. The average correlations for the states are (a) (0.215, 0.354, 0.421, 0.484, 0.642) and (b) (0.228,  0.348, 0.438, 0.543, 0.679) respectively.}
\label{fig:6}
\end{figure*}

\begin{figure*}
            \subfigure[$\epsilon=0$,  Pearson correlation matrices]{\includegraphics[width=8.6cm]
{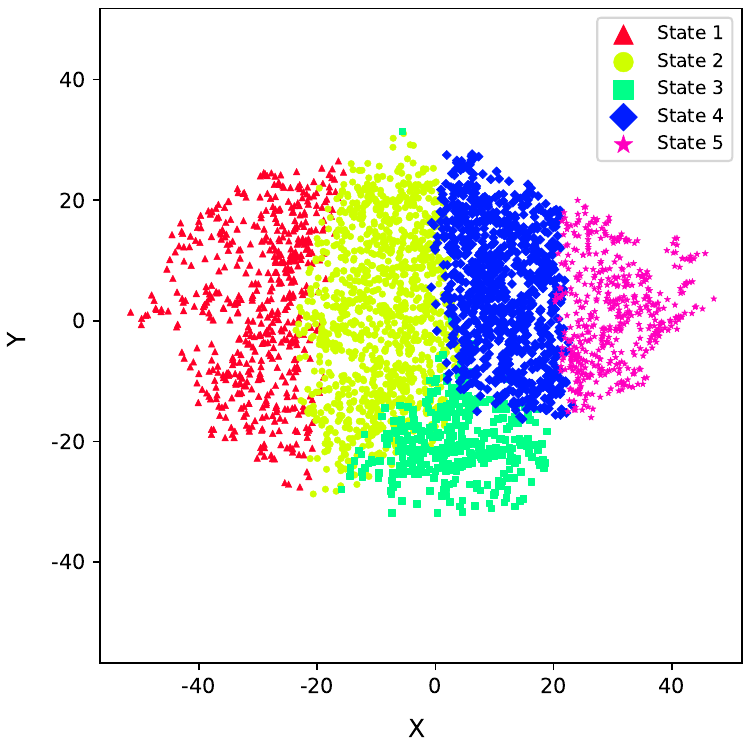}}
            \subfigure[$\epsilon=0$,  sectorial Guhr matrices]{\includegraphics[width=8.6cm]
{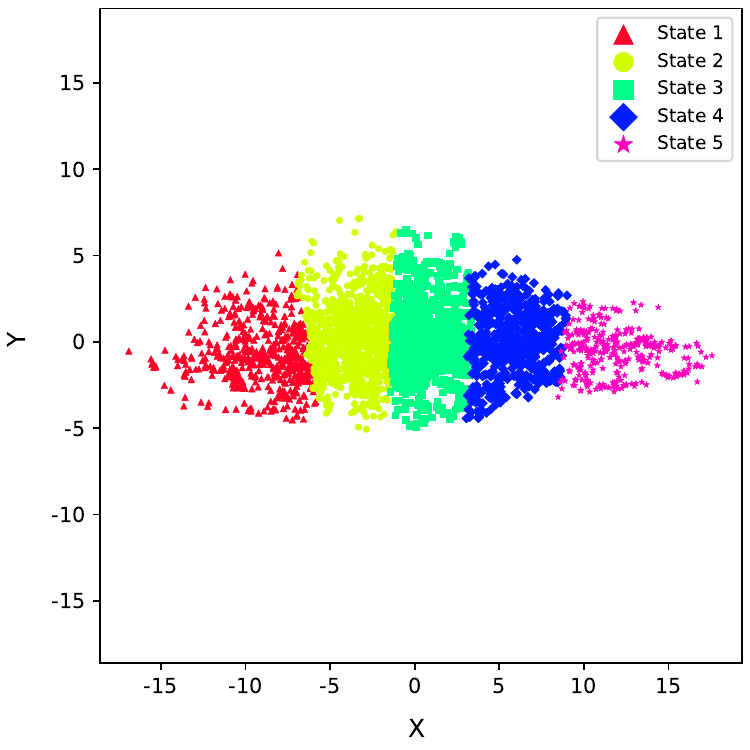}}
 \caption{3D dimensional scaling of 3438 correlation matrices of the five Nikkei 225 market states from January 2006 to December 2019 shown in Fig. \ref{fig:9}: (a) $\epsilon = 0.0$, Pearson correlation matrices and (b) $\epsilon = 0.0$,  sectorial Guhr matrices.}
\label{fig:7}
\end{figure*}

\begin{figure*}[t]
    \centering
    \subfigure[]{\includegraphics[width=8.6cm]{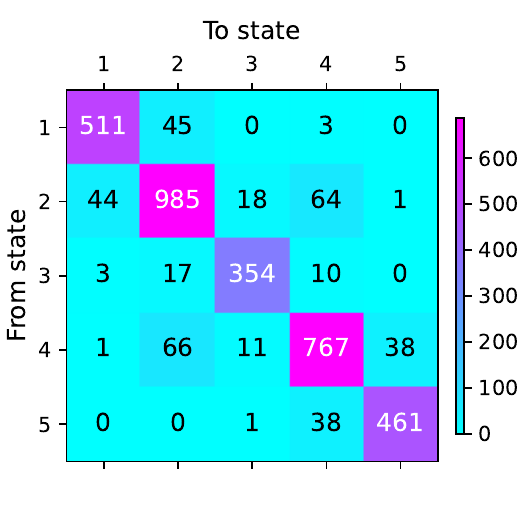}}
    \subfigure[]{\includegraphics[width=8.6cm]{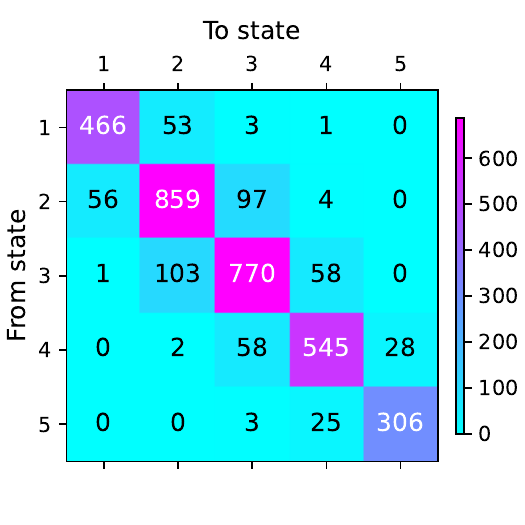}}
  \caption{Transition matrices showing transitions between different market states for Nikkei 225 in Fig.  \ref{fig:9} obtained with (a) Pearson correlation matrices and (b) sectorial Guhr matrices.  The transition matrix corresponding to Guhr matrices is nearly tri-diagonal.   Also, the necessary criterion for Markovianity given in Eq. (2) of \cite{NJP2018} is fulfilled for both.  The equilibrium distributions corresponding to (a) and (b) are $(0.1635, 0.3254, 0.1118, 0.2549, 0.1444)$ and $(0.1529, 0.297 , 0.2699, 0.1834, 0.0968)$ respectively.
}
    \label{fig:8}
\end{figure*}

We perform similar analysis with Nikkei 225 market data and obtain similar results as that of S\&P 500 market data. Figures \ref{fig:6}-\ref{fig:8} show the corresponding results of Nikkei 225 market  for state evolution,  dimensionally scaled matrices projected to two dimensions and transition matrices, respectively. Nevertheless, as seen from Fig. \ref{fig:15} in Appendix \ref{app1}, PM (with optimal $\epsilon$) does not remove market state overlaps even for Pearson correlation matrices. This is in confirmation with previous studies \cite{NJP2018, Springer2019, PhyA24}.  

\section{Conclusions and future outlook}
\label{sec3} 

We have analyzed aspects of complex system dynamics using financial markets as a paradigm.  Using the logarithmic-return-time series of adjusted closing prices with epochs of length 20 trading days shifted by one trading day,  we calculate the Pearson correlation matrices for both S\&P 500 and Nikkei 225 markets.  We obtain the sectorial Guhr matrices by dividing the correlation matrices into sectorial blocks and averaging over intra- and inter-sectorial blocks. We thus obtain $10 \times 10$ and $6 \times 6$ dimensional sectorial Guhr matrices respectively for S\&P 500 and Nikkei 225 markets.  Using these Pearson and Guhr matrices,  we use KM clustering to obtain the market states.  Using these, we analyze the evolution of the market states and the corresponding transition matrices.  We also dimensionally scale Pearson and sectorial Guhr matrices to three dimensions for visualization.  As CG implies averaging, it will suppress noise and indeed the lineal behavior in the market states and quasi tri-diagonality of transition matrices are retained up to a larger number of market states which will likely be useful for certain applications.  Note that there is no guarantee that this will be optimal approach for risk assessment \cite{PhyA24} in view of the fact that we do not use the transition matrix as an additional criterion to optimize $\epsilon$ in potential use of  PM.  We also checked the so-called Guhr covariance method which yield similar but somewhat less convincing results.  

We find that the behavior of the market states for the CG matrices is rather similar to the Pearson correlation matrices. Note that the average correlation which is the dominant feature is almost same for sectorial Guhr matrices and Pearson correlation matrices. In the Appendix \ref{app1}, we discuss the influence of applying PM to the Pearson correlation matrices and find that the resulting Guhr matrices may also have improved transition matrices.  Note that though the features discussed in Ref. \cite{arXiv2023} will disappear if we use PM. The question whether CG affects them or not will have to be analyzed in a future paper as this state exists outside the time horizon of the present paper. Note that the number of relevant variables is reduced to 55 and 21 respectively for S\&P 500 and Nikkei 225 markets thus the making the problem more manageable. It will be interesting to analyze more extreme CG and we plan to reduce sectorial Guhr matrices to $2 \times 2$ and possibly $3 \times 3$ matrices. Preliminary work has shown this to be promising. 

{The method of coarse graining is rather general and applies to other physical systems as well, for example, in analyzing correlations in spin networks \cite{Spn-net} and EEG data \cite{EEG}. Coarse graining over different Hamiltonian blocks with good spin projections in spin networks might lead to new insights but this is for future. Similarly, the multi-band correlation matrix containing intra-frequency band correlations for narrow band filtered EEG signals of healthy subjects during sleep is found to have a close resemblance to Guhr matrices \cite{EEG}.}

\section{Acknowledgements}

The authors are grateful to Anirban Chakraborti,  Thomas Gorin, A. Ra{\'u}l Hern{\'a}ndez Montoya, Christof Jung,  Francois Leyvraz and Thomas H. Seligman for their inputs and suggestions.  P.M.  acknowledges a fellowship from CONAHCYT Project Fronteras 428214.  We gratefully acknowledge financial support from CONAHCYT project Fronteras 10872, CONAHCYT Project Fronteras 425854,  and UNAM-DGAPA PAPIIT AG101122.

\FloatBarrier
\bibliography{CG-refs.bib}

\newpage

\appendix

\setcounter{figure}{0}    

\renewcommand\thefigure{\thesection.\arabic{figure}}
\renewcommand\thesubsection{\thesection.\arabic{subsection}}
\onecolumngrid

\section{Optimization using Power Map on the Pearson correlation matrices}
\label{app1}

In this section, We optimize the number of market states as a function of power map exponent $\epsilon$ \cite{Guhr_2003, Guhr_2010, THS_2013} applied to the Pearson correlation matrices based on the minimum standard deviation of the intra-cluster distances $\sigma_{intra}$ \cite{PhyA24} of the resulting Guhr matrices.  The results are obtained using PM along with CG.  Using the Pearson correlation matrices, we apply the PM and coarse grain these to obtain the corresponding sectorial Guhr matrices. Then, we use the sectorial Guhr matrices to obtain the intra-cluster distances and we choose the parameters (number of clusters and $\epsilon$) minimizing $\sigma_{intra}$.  Same parameters are used for Pearson correlation matrices as well.

\subsection{S\&P 500}

\begin{figure*}[t]
    \centering
\includegraphics[width=12.5cm]{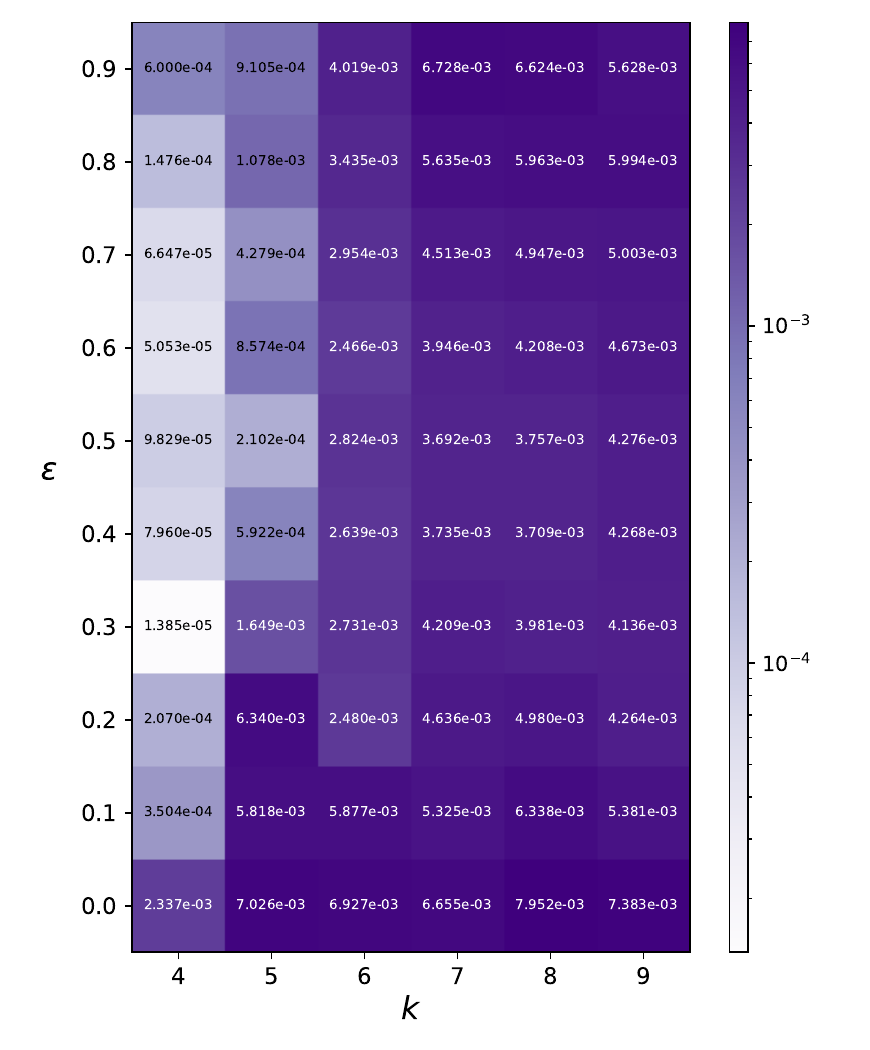}
   \caption{Classification of market states based on the minimum standard deviation of intra-cluster distances $\sigma_{intra}$ for S$\&$P 500 market. We have used 1000 different initial conditions for the KM calculation of $\sigma_{intra}$.  These results are for cases for which we apply PM [Eq. \eqref{4}] to correlation matrices, do the coarse graining and then cluster the Guhr matrices.  For $k \ge 4$, the minima of $\sigma_{intra}$ appears at $k = 5$ and $\epsilon = 0.5$.}
    \label{fig:9}
\end{figure*}

Figure \ref{fig:9} shows the classification of market states based on the minimum standard deviation of intra-cluster distances $\sigma_{intra}$ for S$\&$P 500. We apply PM to Pearson correlation matrices, do CG and then cluster the sectorial Guhr matrices.  Variance $\sigma_{intra}$ is calculated using the standard deviation of intra-cluster measure $d_{intra}$ of 1000 different initial conditions for different values of number of market states (or clusters $k$) and $\epsilon$.  The optimal values (corresponding to minimum $\sigma_{intra}$) for S\&P 500 market with time horizon 2006-2019, for $k \geq 4$,  appears at $k = 5$ and $\epsilon = 0.5$. 

\begin{figure*}
            \subfigure[Pearson correlation matrices]{\includegraphics[width=17.2cm]
{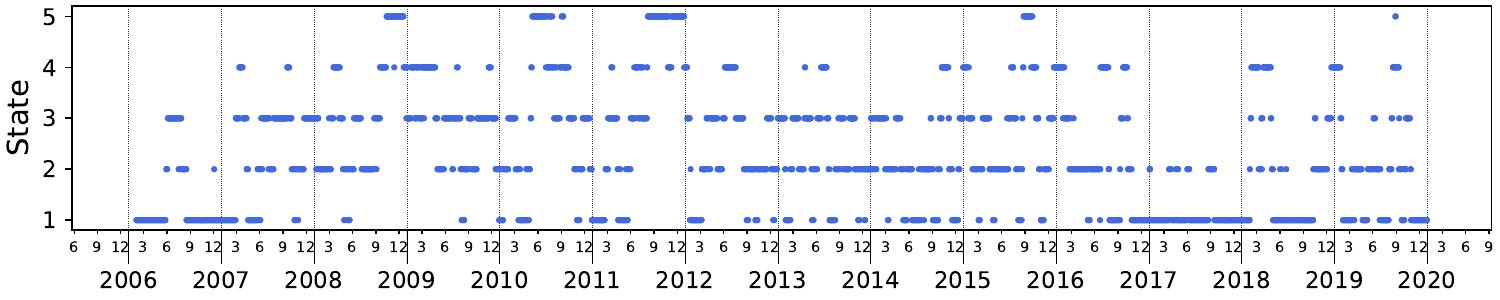}}
            \subfigure[sectorial Guhr matrices]{\includegraphics[width=17.2cm]
{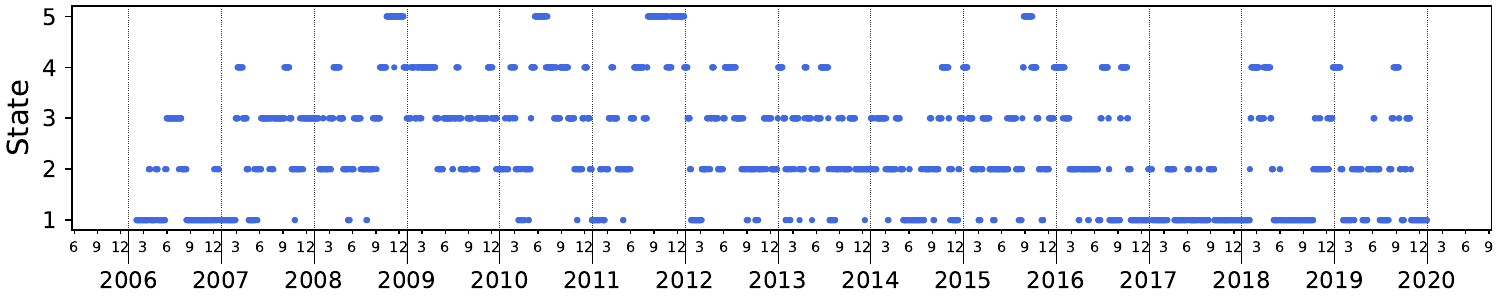}}
 \caption{Time evolution of five market states of the S\&P 500 data using (a) Pearson correlation matrices $C$ and (b) sectorial Guhr matrices $G$.  The state evolution is obtained after performing KM clustering on the 3503 short-time correlation matrices with PM ($\epsilon = 0.5$).  Pearson correlation matrix elements are computed using logarithmic return time series of adjusted closing prices with epochs of length 20 trading days shifted by one trading day. The market states are arranged in order of increasing average correlations. The average correlations for the states are (a) (0.115, 0.208, 0.305, 0.414, 0.563) and (b) (0.116, 0.205, 0.302, 0.411, 0.581),  respectively.}
\label{fig:10}
\end{figure*}

\begin{figure*}
            \subfigure[$\epsilon = 0.5$, Pearson correlation matrices]{\includegraphics[width=8.6cm]
{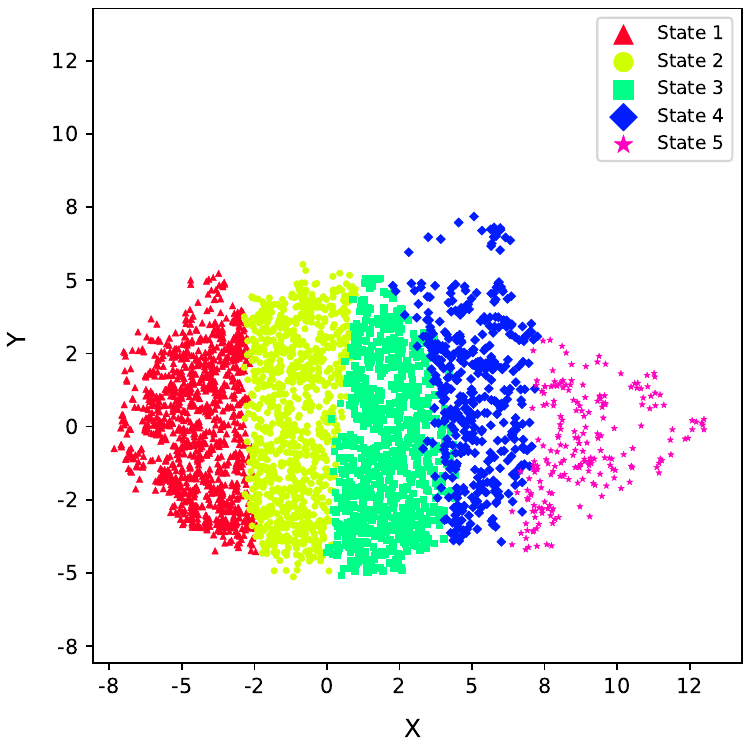}}
            \subfigure[$\epsilon = 0.5$,  sectorial Guhr matrices]{\includegraphics[width=8.6cm]
{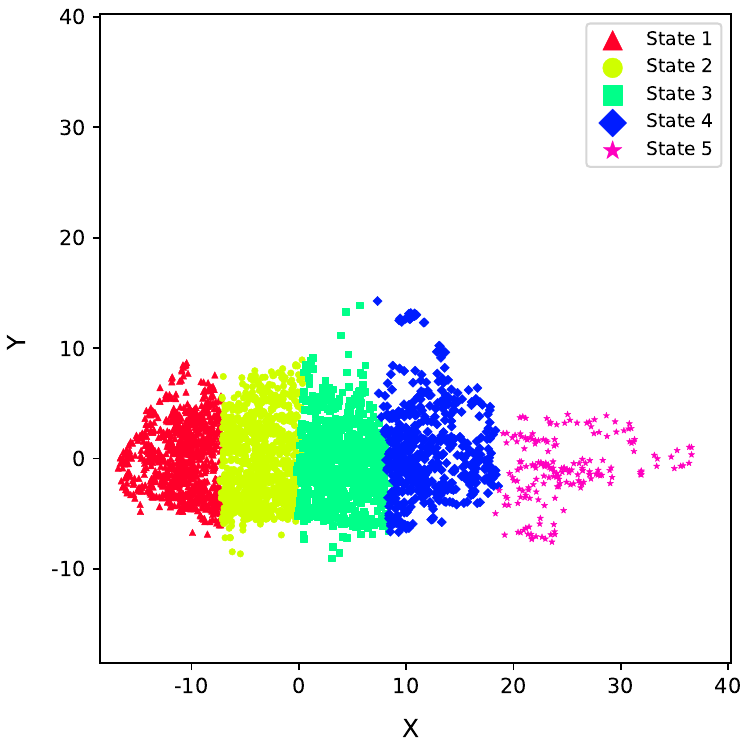}}
 \caption{3D dimensional scaling of 3503 correlation matrices of the five S\&P 500 market states from January 2006 to December 2019 shown in Fig. \ref{fig:6}: (a) $\epsilon = 0.5$, Pearson correlation matrices and (b) $\epsilon = 0.5$,  sectorial Guhr matrices.  Note that the power mapped Pearson correlation matrices behave essentially the same as sectorial Guhr matrices. }
\label{fig:11}
\end{figure*}

\begin{figure*}
           \subfigure[$\epsilon = 0.5$, Pearson correlation matrices]{\includegraphics[width=8.6cm]
{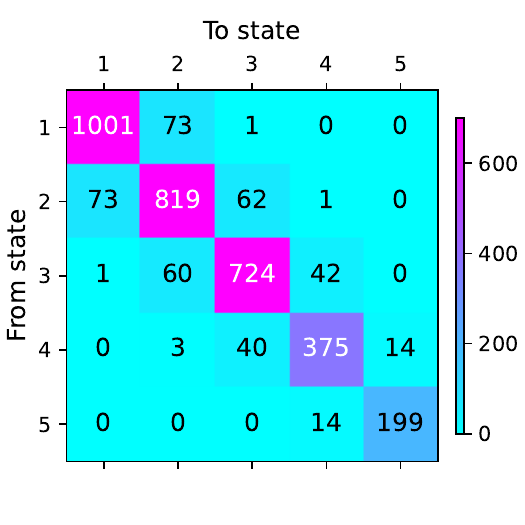}}
            \subfigure[$\epsilon = 0.5$,  sectorial Guhr matrices]{\includegraphics[width=8.6cm]
{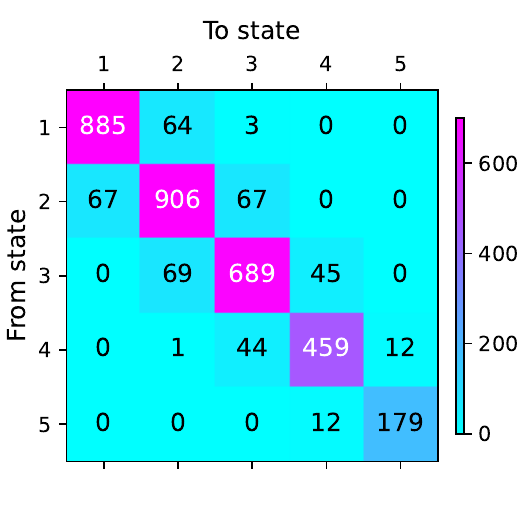}}
  \caption{Transition matrices showing transitions between different market states for S\&P 500 in Fig.  \ref{fig:6} obtained with (a) Pearson correlation matrices and (b) sectorial Guhr matrices.  The transition matrices are near-tridiagonal and the necessary criterion for Markovianity given in Eq. (2) of \cite{NJP2018} is fulfilled, for both. The equilibrium distributions corresponding to (a) and (b) are (0.307, 0.273, 0.236, 0.123, 0.061) and (b) (0.272, 0.297, 0.229, 0.147, 0.055), respectively.}
\label{fig:12}
\end{figure*}

Corresponding to this optimal choice of parameters, the results for state evolution, dimensionally scaled matrices projected to two dimensions and transition matrices are shown in Figs. \ref{fig:10}-\ref{fig:12}, respectively.

\subsection{Nikkei 225}

\begin{figure*}[t]
    \centering
  \includegraphics[width=12.5cm]{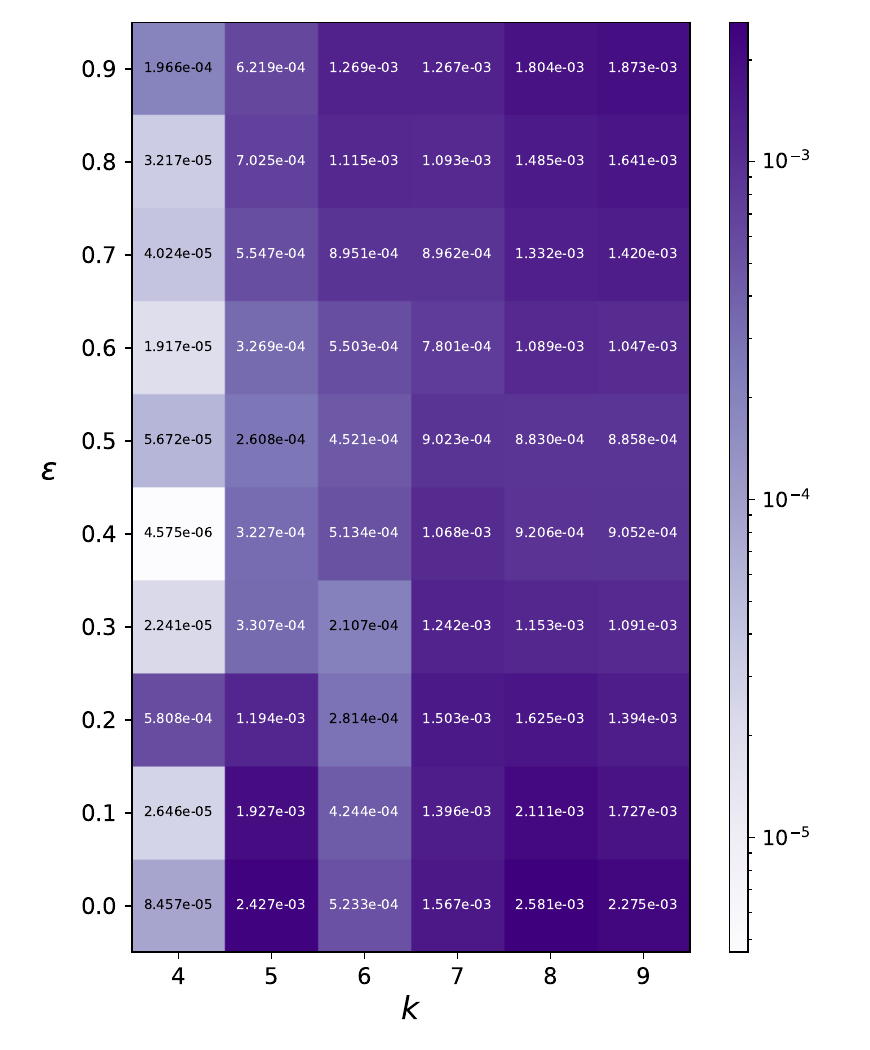}
   \caption{Classification of market states based on the minimum standard deviation of intra-cluster distances $\sigma_{intra}$ for Nikkei 225 market. We have used 1000 different initial conditions for the KM calculation of $\sigma_{intra}$.  These results are for cases for which we apply PM [Eq. \eqref{4}] to correlation matrices, do the coarse graining and then cluster the Guhr matrices. For $k \ge 4$, the minima of $\sigma_{intra}$ appears at $k = 6$ and $\epsilon = 0.3$.}
    \label{fig:13}
\end{figure*}

Figure \ref{fig:13} shows the classification of market states based on the minimum standard deviation of intra-cluster distances for $\sigma_{intra}$ for Nikkei 225. We apply PM to Pearson correlation matrices, do CG and then cluster the sectorial Guhr matrices.  Variance $\sigma_{intra}$ is calculated using the standard deviation of intra-cluster measure $d_{intra}$ of 1000 different initial conditions for different values of number of market states (or clusters $k$) and $\epsilon$.  The optimal values (corresponding to minimum $\sigma_{intra}$) for Nikkei 225 market with time horizon 2006-2019, for $k \ge 4$,  appears at $k = 6$ and $\epsilon = 0.3$. 

\begin{figure*}
              \subfigure[Pearson correlation matrices]{\includegraphics[width=17.2cm]
{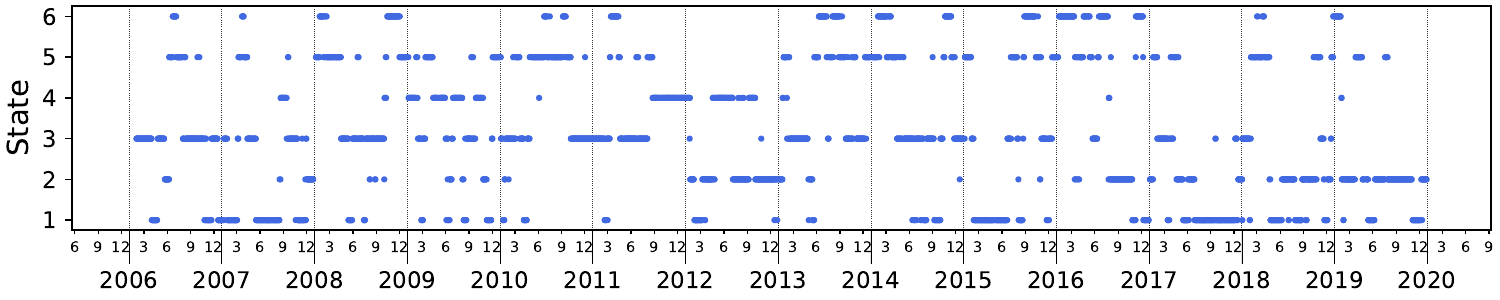}\label{fig:12a}}
            \subfigure[sectorial Guhr matrices]{\includegraphics[width=17.2cm]
{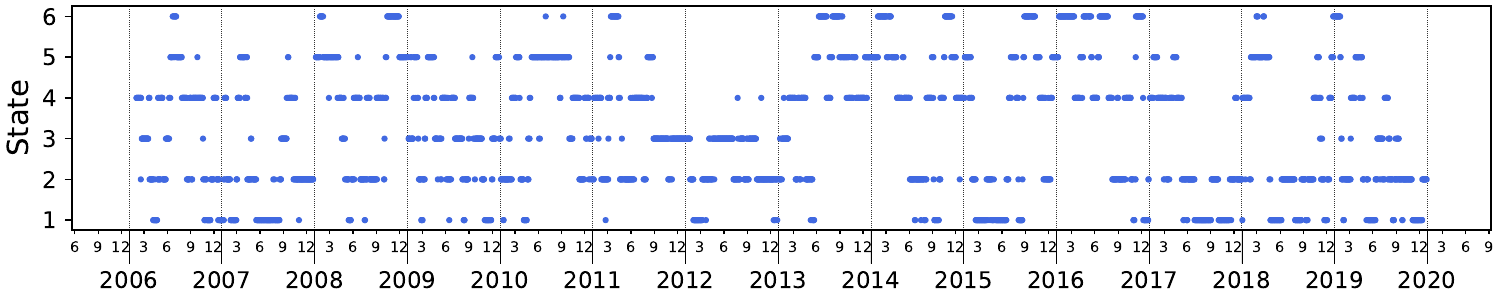}\label{fig:12c}}
 \caption{Time evolution of five market states of the Nikkei 225 data using (a) Pearson correlation matrices $C$ and (b) sectorial Guhr matrices $G$.  The state evolution is obtained after performing KM clustering on the 3438 short-time correlation matrices with PM ($\epsilon = 0.3$).  Pearson correlation matrix elements are computed using logarithmic return time series of adjusted closing prices with epochs of length 20 trading days shifted by one trading day. The market states are arranged in order of increasing average correlations. The average correlations for the states are (a) (0.177, 0.287, 0.324, 0.369, 0.454,  0.600) and (b) (0.178, 0.279, 0.339, 0.375, 0.475, 0.616)  respectively.}
\label{fig:14}
\end{figure*}

\begin{figure*}
            \subfigure[$\epsilon=0.3$, Pearson correlation matrices]{\includegraphics[width=8.6cm]
{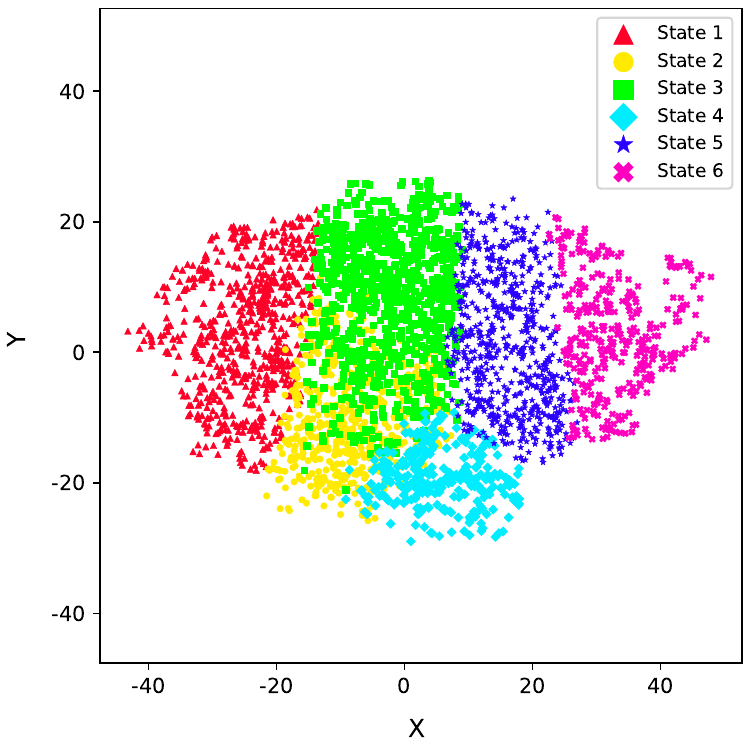}}
            \subfigure[$\epsilon=0.3$,  sectorial Guhr matrices]{\includegraphics[width=8.6cm]
{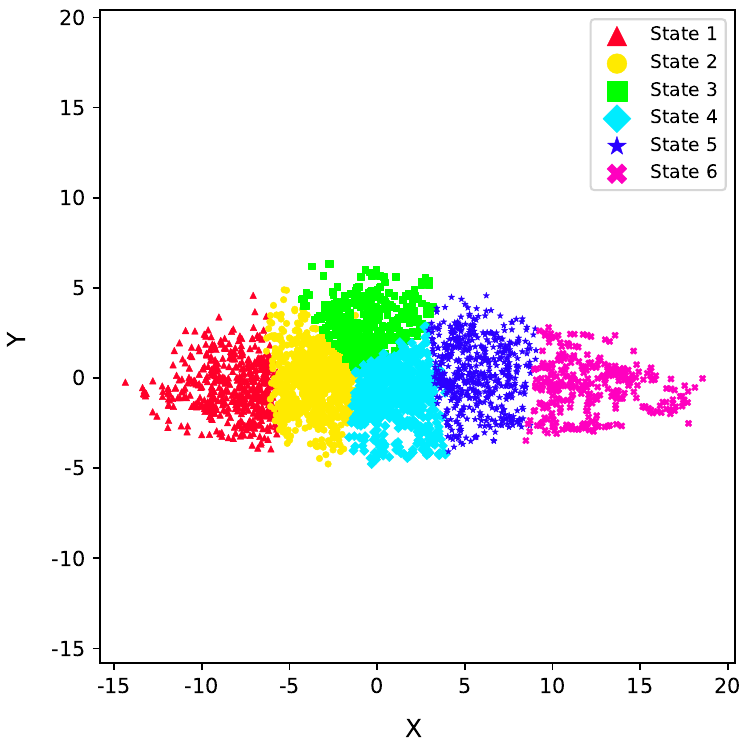}}
 \caption{3D dimensional scaling of 3438 correlation matrices of the five Nikkei 225 market states from January 2006 to December 2019 shown in Fig. \ref{fig:12}: (a)  $\epsilon = 0.3$, Pearson correlation matrices and (b) $\epsilon = 0.3$,  Guhr matrices. Note that in this case, the PM will not establish near-linear behavior for the market states of the Pearson correlation matrix while the figure for sectorial Guhr matrix is not significantly affected.}
\label{fig:15}
\end{figure*}

\begin{figure*}
             \subfigure[$\epsilon=0.3$, Pearson correlation matrices]{\includegraphics[width=8.6cm]
{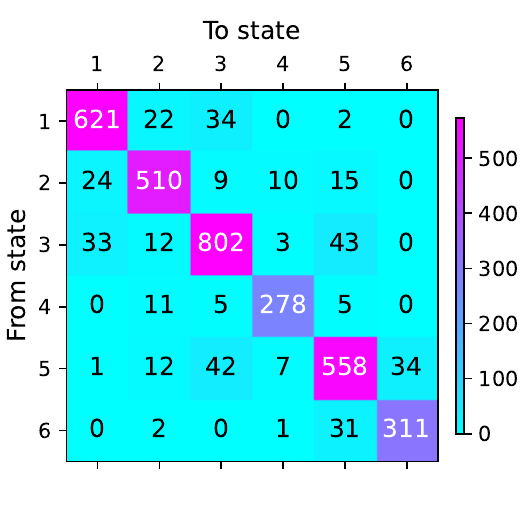}\label{fig:10a}}
            \subfigure[$\epsilon=0.3$, sectorial Guhr matrices]{\includegraphics[width=8.6cm]
{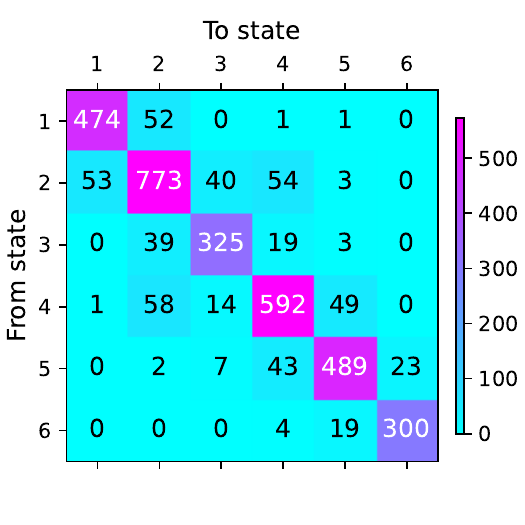}\label{fig:10c}}
 \caption{Transition matrices showing transitions between different market states for Nikkei 225 in Fig.  \ref{fig:12} obtained with (a) Pearson correlation matrices and (b) sectorial Guhr matrices.  The transition matrices are near-tridiagonal and the necessary criterion for Markovianity given in Eq. (2) of \cite{NJP2018} is fulfilled,  for both. The equilibrium distributions corresponding to (a) and (b) are $(0.198,  0.168, 0.257,  0.087,  0.189,  0.100)$ and $(0.155,  0.270,  0.113,  0.206,  0.163,  0.093)$ respectively.}
\label{fig:16}
\end{figure*}

Corresponding to this optimal choice of parameters, the results for state evolution, dimensionally scaled matrices projected to two dimensions and transition matrices are shown in Figs. \ref{fig:14}-\ref{fig:16}, respectively.

\ed